\documentstyle{article}
\begin{document}
\title{Large Scale Weather Control Using Nuclear Reactors}
\author{Moninder Singh Modgil  \\{\it Email:msingh@iitk.ac.in}
\\{\it Department of Physics,} \\{\it Indian Institute of Technology,
Kanpur, 208016},\\ {\it India}}
\date{21 September 2002}
\maketitle

\abstract It is pointed out that controlled release of thermal
energy from fission type nuclear reactors can be used to alter
weather patterns over significantly large geographical regions.
(1) Nuclear heat creates a low pressure region, which can be used
to draw moist air from oceans, onto deserts. (2) Creation of low
pressure zones over oceans using Nuclear heat can lead to
Controlled Cyclone Creation (CCC).(3) Nuclear heat can also be
used to melt glaciers and control water flow in rivers.

\section{ Transforming Deserts}

In this paper we point out that Nuclear reactors can be used for
large scale weather control, e.g., cause rain downpour over
perennial drought areas such as - African Sahara, and create one
of world's most fertile and lush regions. Heat from ground based
nuclear reactors can be used to create a large scale low pressure
region over Sahara, and pull moist air above Atlantic and Arabic
Oceans into Northern Africa. Desert sand is essentially crushed
powdered stone. It is common experience of anyone walking
barefoot on a stone pavement in any large temple of India, that
stone both heats and cools quickly. The stone is almost
unbearably hot for bare feet in summer, and cold in winter. This
is the reason for temperature extremes of both – the unpopulated
deserts, and the densely populated metropolises, which are
essentially forests of stone and concrete.

Sahara air heats up during day and this creates a daily
low-pressure region. The heat from Sahara surface escapes into
outer space during the night, and resulting cooling creates a
high-pressure region, which repels any moist air coming in from
the Atlantic or Arabic Oceans. Thus the suction  effect of low
pressure regions during day heating is nullified. Saharan weather
system is like a spring undergoing small contractions and
expansions – diurnal heating expansions, followed by nocturnal
cooling contractions. Rapid cooling of desert air during night
occurs because of lack of cloud cover, and absence of aerosols in
desert air, which in other circumstances would trap the heat and
prevent sharp temperature drop. The desert weather system is
therefore inherently insufficient to pull moist Atlantic or
Arabic Oceans' air over to interior (non-coastal) regions, and
cause regular rainfall. This is to be contrasted to regular
yearly Monsoon over South Asian Sub-Continent. Green cover in
Indian subcontinent generates aerosols in form of pollen grains,
which help to trap heat, and thus prevent daily highs and lows of
temperature and pressure. This enable formation of sustained low
pressure region, which acts like a pump, to suck moist air from,
Arabic Ocean, Indian Ocean and Bay of Bengal; onto the Indian
Plane. Thus long term Forest Sustained maintainence of a low
pressure region over a large land mass is essential to draw moist
air from an ocean, and cause large scale rainfall, over that
region.

Sustained low pressure over Sahara  could be achieved by –

(1) Heating air by open air ground based nuclear reactors, during
earlier part of night. (2) Injecting aerosols and providing cloud
cover during evenings to reduce loss of heat during night to
outer space.

As the incoming moist air from Atlantic and Arabic Oceans reaches
interior Sahara regions during latter half of night, the nuclear
reactors could be switched off, and the natural nightly cooling
of desert air, combined with presence of artificially injected
aerosols should lead to early morning rainfall. Existing weather
codes running on Super Computers (Crays) in principle should be
capable of modeling and refining this idea. Deserts such as
Sahara (Africa), Gobi (China), Thar (Rajasthan, India), southern
regions of north America (Mexico, Nevada, Texas, California) can
virtually be wiped out from face of this earth. Deserts therefore
are a species threatened with extinction.

The idea can actually be tested in India or Iraq. These countries
already have nuclear reactors. Important issue in such a
experiment is the height above the ground. Lets see what happens
when an open air nuclear reactor is installed on desert plane.
The reactors should be directly exposed to air, without any
concrete building or housing, and switched on. As the heat
content builds up, the ensuing local low pressure will start
sucking air from neighboring regions. This will create a desert
sand-storm. We see that now the energy of reactor is being used
up in accelerating the sand particles. This means that a large
part of the energy will not be available, to suck in moist air
above oceans. We now also see that while energy from sun, does
generate strong winds in desert, this energy is being used in
accelerating sand particles in air, and is not getting used in
bringing in moist ocean air. The sand thus is damping out the
circulation being generated by heating of ground by sun, and
heating of air close to ground by nuclear reactor.

Now I introduce reader to concept of boundary layer. When wind
blows over ground, energy of wind is transferred to sand or dust
particles on the ground, and sends them flying into air. This is
really a form of friction between the stationary ground and
moving air (wind). The region of wind near the ground therefore
is moving slower because it is in the so called 'boundary layer',
region of air slowed by friction from ground or boundary. If one
glances up on a windy cloudy day, one notices, that clouds high
up are moving faster while clouds closer to ground are moving
slower or hardly moving. The reason is that closer clouds are in
boundary layer. Therefore it is important to keep the reactor out
of boundary layer, so that energy transferred from reactor to
wind, does not get dissipated in sending sand spinning into air.
What if the reactor is at a height of say 50 or 100 meters above
ground? It can be installed on top of a TV tower or on roof a
skyscraper. Now the winds being circulated will be generated 50
or 100 meters above ground, and will not dissipate all their
energy in sucking sand particles from desert floor. However, a
small quantity of lighter sand or dust particles will still be
sucked into air, and sent high into atmosphere (1 to 2
Kilometers). We will have a large scale circulation build up in
desert air, which will not be sand strom. These lighter and
smaller dust  particles sucked out from ground and sent spinning
high into the atmosphere will act as natural aerosols - centers
for formation of rain droplets.

Timed controlled use of nuclear reactors and Aerosol injections
can be used to provide daily cloud cover and thus cool hot Indian
Metropolises such as Delhi in summer, and reduce mass suffering
of populace.

\section{ Melting Glaciers – Military and Chess in Indian
Sub-Continent}

Weather control using Nuclear Reactors and Cray Super Computers;
also has potential military applications. Its possible to use
heat generated by nuclear reactors to control melt down of snow
in Himalayan glaciers. These glaciers feed rivers, which run
through plains of Pakistan, India, Bangladesh and China. Thus, in
principle, an enemy with nuclear power, and a good Intelligence,
can install nuclear reactors in Himalayan Glaciers, and use them
for disrupting economies of these nations. The whole process
would look like a series of natural catastrophes. This Indirect
Nuclear Threat (INT) has a very good chance of success in
destroying an enemy, in contrast to the direct nuclear threat,
which has only deterrence or preventive capability via MAD -
Mutually Assured Destruction. Vital question for countries in
this region is  - has a foreign power been actively pursuing this
particular method of Indirect Nuclear Threat (INT)?

 Lets consider the following issue of national and international
  interest, as a possible motive for application of this idea.
" Will the Western Capitalist Conglomerate really need Kashmir as
Part of a Long Term Strategic Plan to overthrow the remaining
twin threats Communism in China and Muslim fundamentalists in
Pakistan? "

Precedents:

To support analysis of this issue, we cite in beginning following
precedents.

(1) Economic breakdown of USSR (Communist Russia) was used to
dismember the communist state. The issue whether the economic
breakdown was a natural result of communist policies or caused by
capitalist intelligence does not concern us in this paper.

(2) In Kosovo, plea of genocide lead to military intervention of
UNO and physical occupation of Kosovo by NATO troops.

(3) India was befriended and latter deceived by China, and lost
strategically important Himalayan peaks.

The first precedent tells us that economic breakdown can be used
as an effective weapon for battling or neutralizing threat of an
enemy nation. The second precedent tells us that western
capitalism has perfected and demonstrated effectiveness of a new
tactical weapon, which exploits weaknesses of local people of a
geographical region. The new tactical weapon is –

"Combined use of international diplomacy (UNO), military strategy
and technological know how, over a period of few years."

 This eventually leads to the physical occupation of the
  geographical region by troops of western Capitalist nations, on
 humanitarian grounds. One notes that in Kosovo, it was the Muslim
 population, which was getting massacred, and the Christian
 Capitalists intervened as their saviors. A similar genocide of
  Muslims in Kashmir is likely to lead to a repeat – UNO occupation
  of Kashmir. It appears that the Indian Nation may be heading for
  another deception – by USA this time.  India may loose rest of
  Kashmir to USA. We note with alarm that -
USA already has the resource of a large number of existing spare
nuclear reactors. These existing spare nuclear reactors were the
power source of their older fleet of 'nuclear propelled
submarines'.

Its likely that this idea of using Indirect Nuclear Threat,
emerged, as a result of brain storming on how to use or dispose
of these old reactors. It's extremely likely, that the idea has
already been successfully tested, in causing floods in Pakistan
few years' back. Support for such a thesis is provided by
following reports –

(1) Western civilians were airlifted to the Siachin Glacier
region using helicopters by Pakistanis. Their purpose or motive
was not clear. This happened just when the Pakistanis first moved
into the region.

(2) Pakistani authorities captured smuggled radioactive nuclear
fuel on Afghan-Pakistan border. Reasons for smuggling were not
clear.

One can visualize that having demonstrated the success of this
idea; the USA intelligence would have shifted to using newer and
smaller, state of the art, nuclear reactors. Its likely that
these new state of art small nuclear reactors are currently being
deployed in Himalayan Glaciers to weaken or finish off the
remaining fortresses of threats to western capitalist society –
Muslim Pakistan, and Communist China. Flooding of BrahmaPutra
would effect Tibet, China and Bangladesh, while flooding of
Sindhu would effect Pakistan. Flooding of Ganges and Yamuna can
be used to weaken opposition to Christianity in Indian states of
Uttar Pradesh and Bihar, and as a general tactical weapon.

On hindsight the communist Chinese military planners did well for
themselves, to capture the Himalayan peaks in early sixties. The
reason being that this geographical region contains glaciers and
snow, which feeds the rivers, which flows through their planes.
Strategically, it was a good move on part of Chinese, even though
the idea of indirect nuclear threat (INT) may not even have
occurred to the Chinese military planners.

 In chess, as in warfare and life – good strategic moves are
 moves designed to avoid loss through a short tactical combination.
 In any general conflict, such as war, chess etc., knowledge
  of opponent's plan can be used to neutralize the offensive.
   This is the reason for existence of intelligence agencies.
    However, if an opponent is given time to carry out a plan,
     due to lack of awareness on one's part, defeat becomes
     unavoidable. An ambitious chess player studies his and
     other players' past losses to learn lessons for future
     games. It has been estimated that to become a chess
     grandmaster, one has to loose 500 tournament chess games.
     Aside from the strategic use of Himalayas – Indirect Nuclear
     Threat. (INT) mentioned in this paper, its possible to
     tactically use the height advantage, with space-age weapons.
     USA may be having long terms plans for colonizing rest of
     Himalayas and use as a powerful platform for exercising
     world control.

\section{Generating Cyclones using Nuclear Reactors -
Storm-Bringer Cometh}

Its even possible to generate severe cyclones using a nuclear
reactor about 30 meters under the water in a sea such as Bay of
Bengal. The nuclear reactors heat up the water, which in turn
warms the air. Energy equivalent of many nuclear bombs can be
injected into the atmosphere. The warmth in the atmosphere
creates a local low-pressure region, and cooler air from
neighboring areas starts moving towards it and assumes the form
of a cyclone. Its possible to analyze how the resulting cyclone
will move, using the Cray Super computers USA has for weather
prediction. In an appropriate weather condition, and wind
directions in Bay of Bengal, the method of Indirect nuclear
threat (INT) may have actually been used to 'create the Orissa
cyclone towards end of 1999'.  The motive of the cyclone could be
conversions to Christianity. One notes with concern the preceding
Killing of the Australian missionary by so called Dara Singh in
the region.

Also note that UNO had declared the 1990-2000 decade, as the
decade of natural disasters. It appears to be part of propaganda
of a long plan, already in an advanced stage of completion.

\section{Melting Polar Ice Caps using Nuclear Reactors – Global
Flooding and Green House Brain Wash}

There have been reports of thinning of ice cover in Arctic (North
Pole). It has been attributed to 'Green House' effect. It has
been estimated that the North Pole's ice cap will be totally
melted within next 50 years. The area will then be available for
mining of oil deposits and other valuables. Its likely that the
melting of North polar ice cap is being accomplished by nuclear
reactors embedded within ice cap. It's likely that research on
'Green house' effects, is part of propaganda to distract mass
attention from actual causes. It's likely that Western Capitalist
Oil and other mining corporations sponsor such a project. We may
use the short hand 'Greenwash', for 'Green house brain wash' –
inspired by the 'green' color of US currency - 'dollar'.

\section{Conclusions}

It was Einstein who first pointed out that matter was a hidden
source of enormous energy, in the sense that matter could be
converted into energy. Since almost the whole mass of an atom of
matter resides in its nucleus, therefore this form of energy  may
be termed as 'nuclear energy'. Einstein suffered a nervous
breakdown when atomic energy was unleashed in a military
application over Japan. Einstein had visited Japan in 1920s and
liked the nation. His psychological suffering lasted for a period
of nearly 5 years. Here we see first instance of successful
application of nuclear energy – 'how to drive Einstein mad'.
Other applications of Atomic energy to date include, nuclear
power plants, nuclear propelled submarines and ships, and last
but not the least, nuclear medicine. Modern Medical techniques of
MRI (Magnetic Resonance Imaging), PET (Positron Emission
Tomography) scan, are essentially medical applications of nuclear
energy. Einstein's medical condition was possibly triggered as a
'Karmic back reaction', of hundreds of thousands who suffered in
the nuclear blasts over Hiroshima and Nagasaki. The fact that
those who actually delivered the bomb did not suffer just
indicates their minor role in chain of events. It appears that
there may be a connection between nuclear energy and one's
psychological state, which needs to be explored.

In this paper we have outlined large-scale weather control as
another potential application of nuclear energy. The method is
simplicity itself – use the heat from a nuclear reactor to melt
ice, or heat air and water. We observe that in these
applications, the nuclear energy is doing the job of sun. One is
reminded of Oppenheimer's experience of witnessing the first
man-made nuclear explosion, 'brighter than a thousand suns'. It
would be worthwhile for world's government to see if such a
mischief is being carried out unannounced in their backyards. It
would be also worthwhile to explore if 'thinning Arctic ice
cover' and 'Global warming' is a play of 'few mischievous
capitalist boys'.
\\
In conclusion we cite the hit lead song from the film refugee
which was partly instrumental in leading the author to these
ideas. The film is about two lovebirds, separated by a man-made
border. The lovers live near a river, which flows, through India
into Pakistan. The first line of the song is as follows.
\\Panchi, Nadia, Pavan ke Jhonke; koi sarhad na  inhein roke.
\\
Sarhad Insanoo ke liye hei; Soocho, tumne aur maine kya paya
Insaan hoke.

The translation reads -
\\No boundary stops Bird, River, and gusts of wind.
\\Boundaries are for humans; think, what you and I gained, by being human.

\end{document}